\documentclass{aastex7}

\accepted{4/11/2025 for AJ}

\shorttitle{TESS Light Curve of W Serpentis}
\shortauthors{Gies et al.}

\begin{document}

\title{TESS Light Curve of the Interacting Binary W Serpentis}

\correspondingauthor{Douglas Gies}
\email{dgies@gsu.edu}

\author[0000-0001-8537-3583]{Douglas R. Gies}
\affiliation{Center for High Angular Resolution Astronomy and 
Department of Physics and Astronomy, Georgia State University, 
P.O. Box 5060, Atlanta, GA 30302-5060, USA} 
\email{dgies@gsu.edu}

\author[0000-0003-2075-5227]{Katherine A. Shepard}
\affiliation{Physics and Astronomy Department, Vanderbilt University, 
PMB 401807, 2301 Vanderbilt Place, Nashville, TN 37240, USA}
\email{katherine.shepard.1@vanderbilt.edu}

\author[0000-0002-9811-5521]{Aman Kar}
\affiliation{Department of Physics and Astronomy, Georgia State University, 
P.O. Box 5060, Atlanta, GA 30302-5060, USA} 
\altaffiliation{RECONS Institute, Chambersburg, PA 17201, USA} 
\email{akar5@gsu.edu}

\author[0000-0002-2806-9339]{Noel D. Richardson}
\affiliation{Department of Physics and Astronomy, Embry-Riddle Aeronautical University, 
3700 Willow Creek Road, Prescott, AZ 86301, USA}
\email{richan11@erau.edu}

\begin{abstract}
The unusual light curve of the massive eclipsing binary W~Ser was recently observed 
with high S/N and fast cadence by the NASA TESS mission. The TESS light curve 
records two eclipses and relatively fast variations outside of the eclipses. 
The eclipse timings verify the period increase of the binary, and the period derivative 
implies a mass transfer rate in excess of $10^{-5}$ $M_\odot$ yr$^{-1}$. 
The light curve shows a fading trend from just after an eclipse until the 
start of the next eclipse.  The brightest flux source in the system is the 
accretion torus surrounding the mass gainer star, and we argue that these 
orbital-phase related fadings are the result of the injection of cooler gas 
from the mass donor entering the outskirts of the accretion torus. 
There are cyclic variations in the out-of-eclipse sections of the light curve
that vary on a 2.8 day timescale.  This equals the orbital period for gas in 
the outer regions of the accretion torus, so the photometric variations are 
probably the result of transitory, over-dense regions that form at the rim
of the accretion torus. 
\end{abstract}



\section{Introduction} 
\label{sec:introduction}

Massive stars are often found in close pairs that are destined to become interacting binaries \citep{deMink2013}. 
The mass transfer and loss that occurs during their interacting stage will dramatically alter their evolutionary paths.  
The initially more massive star will be the first to grow
in size to fill its Roche radius.  This mass donor star will then begin mass transfer to the gainer companion at an  
ever increasing rate as the separation decreases.  Once the mass ratio reverses, the system will enlarge  
with continued mass transfer at a decreasing rate.  The mass donor will eventually lose most of its
outer envelope, while the mass gainer will become a massive rapid rotator.  Eventually the gainer will attain  
critical rotation and will no longer be able to gather mass from its surrounding accretion disk.  The accretion 
disk will become a holding-zone torus that will leak gas away from the binary through the outer L3 Lagrangian
point to create an equatorial circumbinary outflow \citep{Lu2023}.  How much mass is lost at this stage is critical to 
the final outcome of the binary system.  

This stage of intense mass transfer is very short ($\approx 10^4$ years; \citealt{Wellstein2001,vanRensbergen2008}), 
so binaries caught at this evolutionary point will be rare.  Nevertheless, there is a class of Algol-type eclipsing binaries that are 
probably experiencing this phase of strong mass transfer.  These are members of the W Serpentis category of binary stars that 
show abundant evidence of mass transfer and an optically thick torus that blocks out the gainer star from view.
There are some dozen known examples of which the brightest (and most famous) is probably $\beta$~Lyrae \citep{Mourard2018}. 
W~Ser (HD 166126, TIC 106163940) is the prototype of the category. It is an eclipsing binary with an orbital period of about $14.17$ days 
\citep{Guinan1989,Erdem2014,Shepard2024}. The visible spectrum is dominated by deep shell lines 
and strong emission lines that mainly originate in a thick circumbinary disk \citep{Bauer1945,Sahade1957,Barba1993}. 
However, the ultraviolet spectrum shows {\bf an} extended, short wavelength continuum and an array of highly excited 
emission lines that implicate a hot temperature for the hidden mass gainer \citep{Plavec1989,Weiland1995,Sanad2013}.  
New echelle spectroscopy of W~Ser has revealed for the first time the faint spectral absorption lines 
associated with the cool mass donor star \citep{Shepard2024}.  The donor star's radial velocity semiamplitude can be compared to 
the star's projected rotational velocity to determine the system mass ratio by assuming that the donor fills its 
Roche lobe.  The resulting mass estimates are 2.0 and $5.7 M_\odot$ for the donor and gainer, respectively \citep{Shepard2024}.
These results indicate that W~Ser is a massive binary that is experiencing an extreme mass transfer rate and systemic mass loss. 

The light curve of W~Ser presents a number of puzzling features that are described by \citet{Guinan1989}.  
Timings of the deep primary eclipses from over the  last century indicate that the 
orbital period is increasing at a rate of 19 seconds per year, which implies a large
mass transfer rate \citep{Erdem2014}.  However, the secondary eclipse has eluded detection presumably because the 
mass donor star is quite cool compared to the mass gainer and its accretion torus \citep{Shepard2024}.  
\citet{Shepard2024} present a discussion of the All Sky Automated Survey (ASAS) light curve \citep{Pojmanski2004} 
that was obtained between 2001 and 2004.  It shows an unusual jump in brightness from before to after the primary eclipse 
as well as relatively large $\pm 0.25$ mag stochastic variations that are unrelated to orbital phase.  
We can now explore the origin of these light curve properties thanks to measurements made by the NASA TESS 
mission \citep{Ricker2015} in 2024 July.  Here we present an analysis of the new TESS light curve of W~Ser to investigate
the long-term, orbital, and fast variations that are clearly revealed for the first time. 


\section{Ephemeris and Period Change}
\label{sec:ephemeris}

The NASA TESS mission obtained its first light curve of W~Ser over 26 days in Sector 80 at a cadencce of 200 seconds.
The TESS spectral throughput corresponds to a band covering the 6000 -- 10000 \AA\ range. 
The TESS pixel scale is approximately 21 arcsec per pixel, so blending with the flux of other nearby stars 
can present a problem for data analysis.  However, an inspection of the Gaia DR3 catalog \citep{Gaia2023} 
indicates that the combined flux of all the other sources within a few pixels of the target amounts to no more 
than $1\%$ of the flux of W~Ser, so no correction was made in the light curve model for additional third light (Section 3).
We obtained the MIT Quick-Look Pipeline version of the light curve \citep{Huang2020} from the 
Mikulski Archive for Space Telescopes\footnote{https://mast.stsci.edu/portal/Mashup/Clients/Mast/Portal.html}. 
We present here the systematics-corrected flux that is formed by applying small instrumental corrections without detrending. 
We removed those measurements that had bad calibration quality flags.
The normalized flux as a function of time is plotted in Figure~1.  
The TESS observations recorded two deep, primary eclipses 
and the faster variations seen outside of the eclipses.

\begin{figure*} 
        \centering
	\includegraphics[width=19cm]{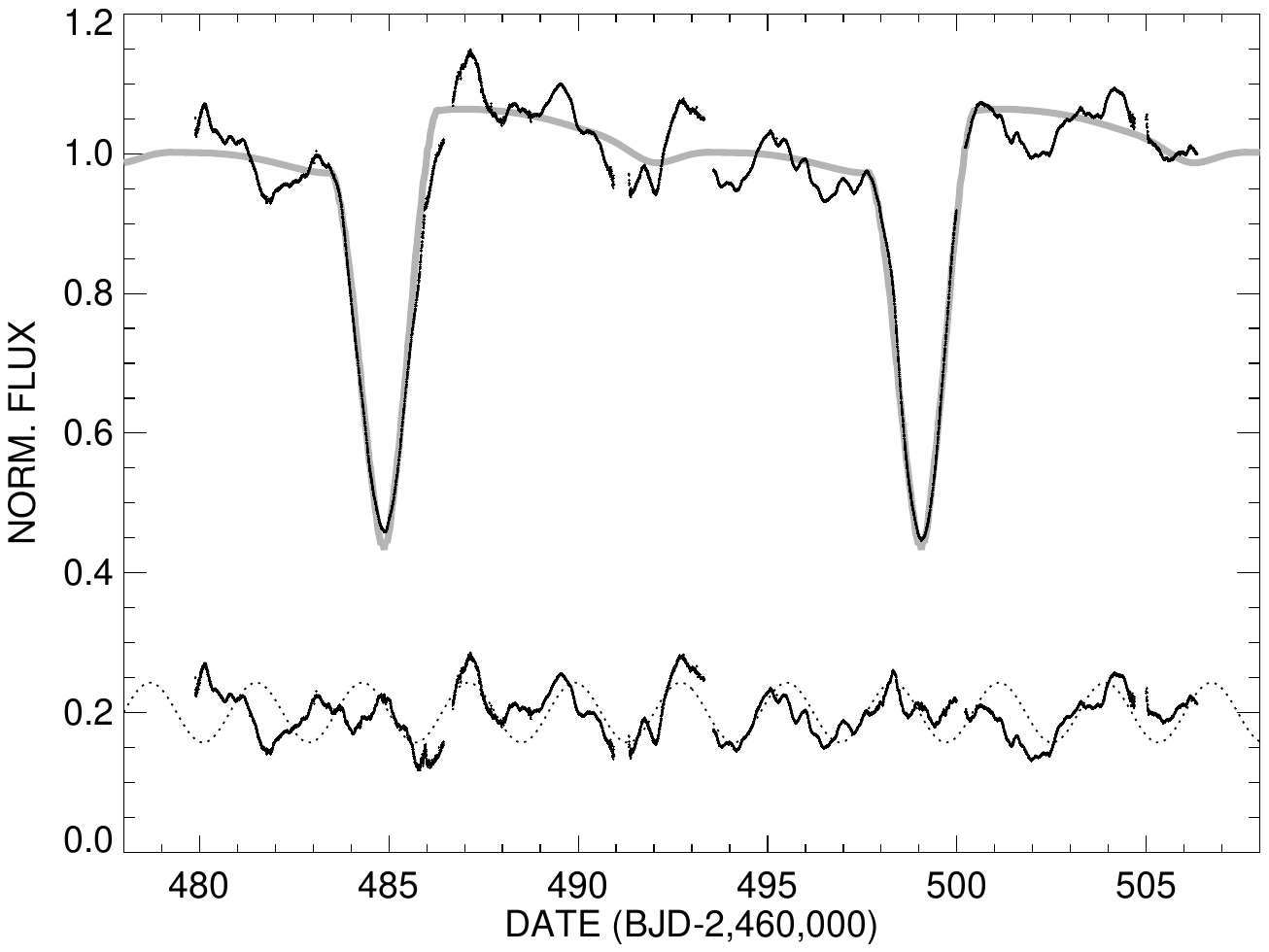}
	\vspace{-1.5cm} 
	\caption{The TESS light curve of W~Ser from 2024 July shown as  
        normalized flux as a function of barycentric Julian date (tiny black symbols). 
        The grey-shaded line shows the ELC model light curve that includes a 
        systematic temperature drop along the rim of the accretion torus surrounding 
        the mass gainer.  The residuals from the fit are shown below (offset by $+0.2$) 
        together with the sinusoidal signal from the peak in the power spectrum (dotted line).}
        \label{fig:light-curve}
\end{figure*}

The excellent quality of the light curve reveals subtle asymmetries in the shapes of the eclipses, 
so fits of eclipse mid-point depend upon the time range included in the fit.  We made parabolic 
fits to the lower parts of the eclipse plot over samples from $50\%$ to $10\%$ of the eclipse depth. 
We then constructed a linear fit of eclipse time as a function of residual depth, and we extrapolated to zero 
depth to derive the estimated eclipse time.  The eclipses occurred at BJD 2,460,484.8759 $\pm 0.0046$ and 
2,460,499.0616 $\pm 0.0046$, where the uncertainties are taken conservatively as the standard deviation 
of the individual fits at different line depths.  

The difference in the eclipse times yields a period of $P = 14.1857 \pm 0.0065$ at the epoch
of the TESS observations.  \citet{Erdem2014} found a net period increase of $18.8 \pm 0.4$ 
seconds per year from the historical record of eclipse timings, and their predicted 
period at the TESS epoch of $P = 14.1845 \pm 0.0090$ agrees within uncertainties with that 
derived from the TESS observations.  

The period derivative is related to the mass transfer rate in the case of conservative 
mass transfer \citep{Erdem2014}.  Comparing the current period from the TESS light curve 
with the results from  \citet{Erdem2014} yields  a period increase rate of 19.7 seconds per year.
Adopting the masses from \citet{Shepard2024}, 
this implies a mass transfer rate of $1.7 \times 10^{-5} M_\odot$ yr$^{-1}$.  
This estimate is a lower limit if some fraction of the mass transfer is lost 
through the outer Lagrangian point \citep{Hilditch2001}.  


\section{Light Curve Model}
\label{sec:model}

\citet{Shepard2024} described a fit of the ASAS light curve that they calculated using 
the ELC binary star code \citep{Orosz2000}.  The model creates an orbital light curve 
based upon the flux from an optically thick torus surrounding the mass gainer star and 
from the fainter mass donor star.  The geometry of components is shown in Figure~2. 
We adopted the same basic parameters to make a fit of the orbital variations in the 
TESS light curve.  These are summarized in Table~4 of the paper by \citet{Shepard2024}. 
The torus extends outwards from the gainer star to a radius of $14.9 R_\odot$ at an 
opening angle of $\pm 25^\circ$ from the orbital plane.  The gas temperature in the 
torus decreases from 14.3~kK at the inner boundary to 8.0~kK at the disk rim, all 
much higher than the temperature of the mass donor $T_{\rm eff} = 5$~kK.  The vertical 
extent at the gas torus rim ($\pm 6.9 R_\odot$) is large enough to completely block the 
flux from the gainer star for an orbital inclination of $i=101^\circ$ (Fig.~2).  

\begin{figure*} 
        \centering
	\includegraphics[width=15cm]{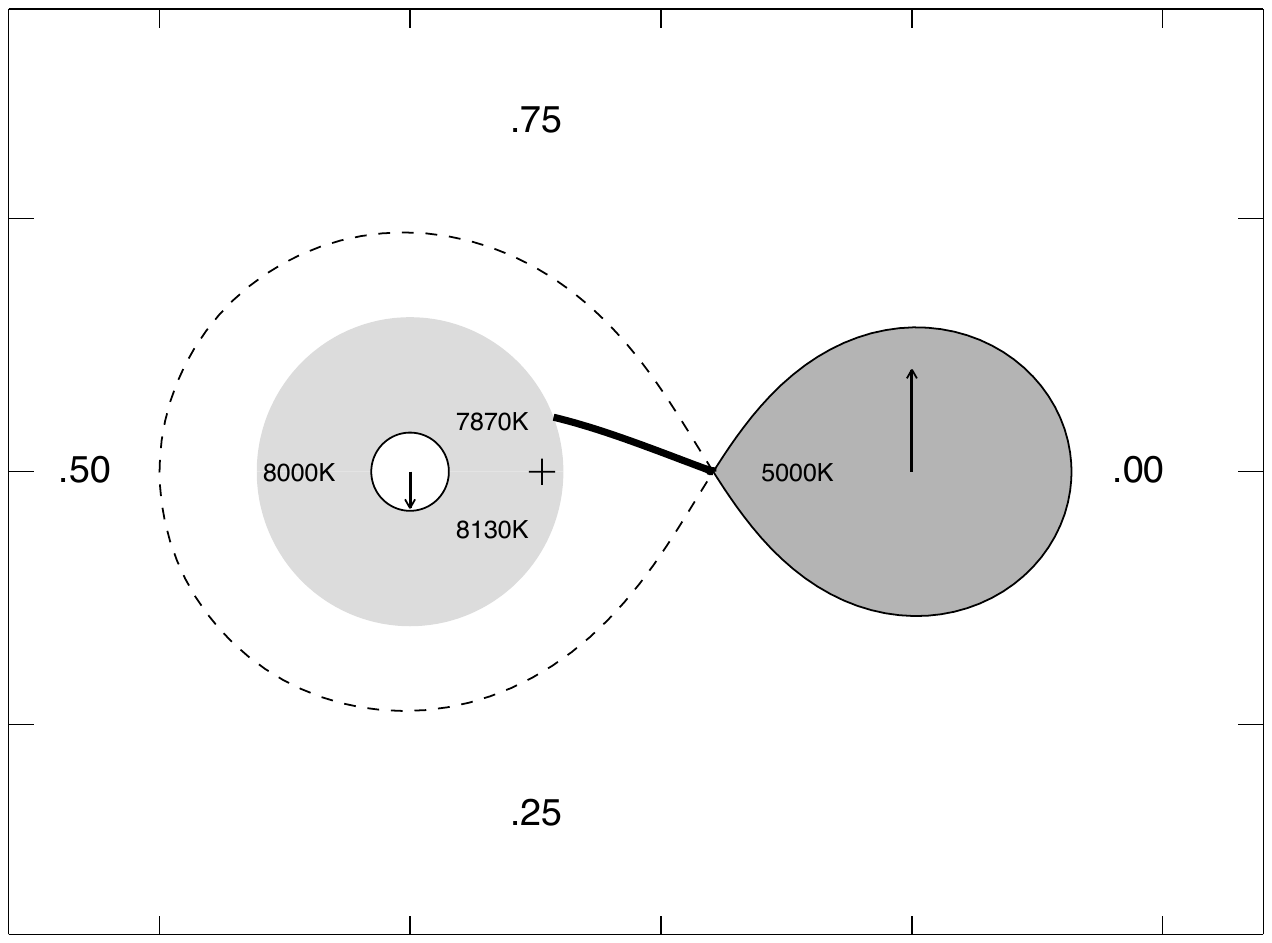} \\
        \vspace{-2.5cm} 
	\includegraphics[width=15cm]{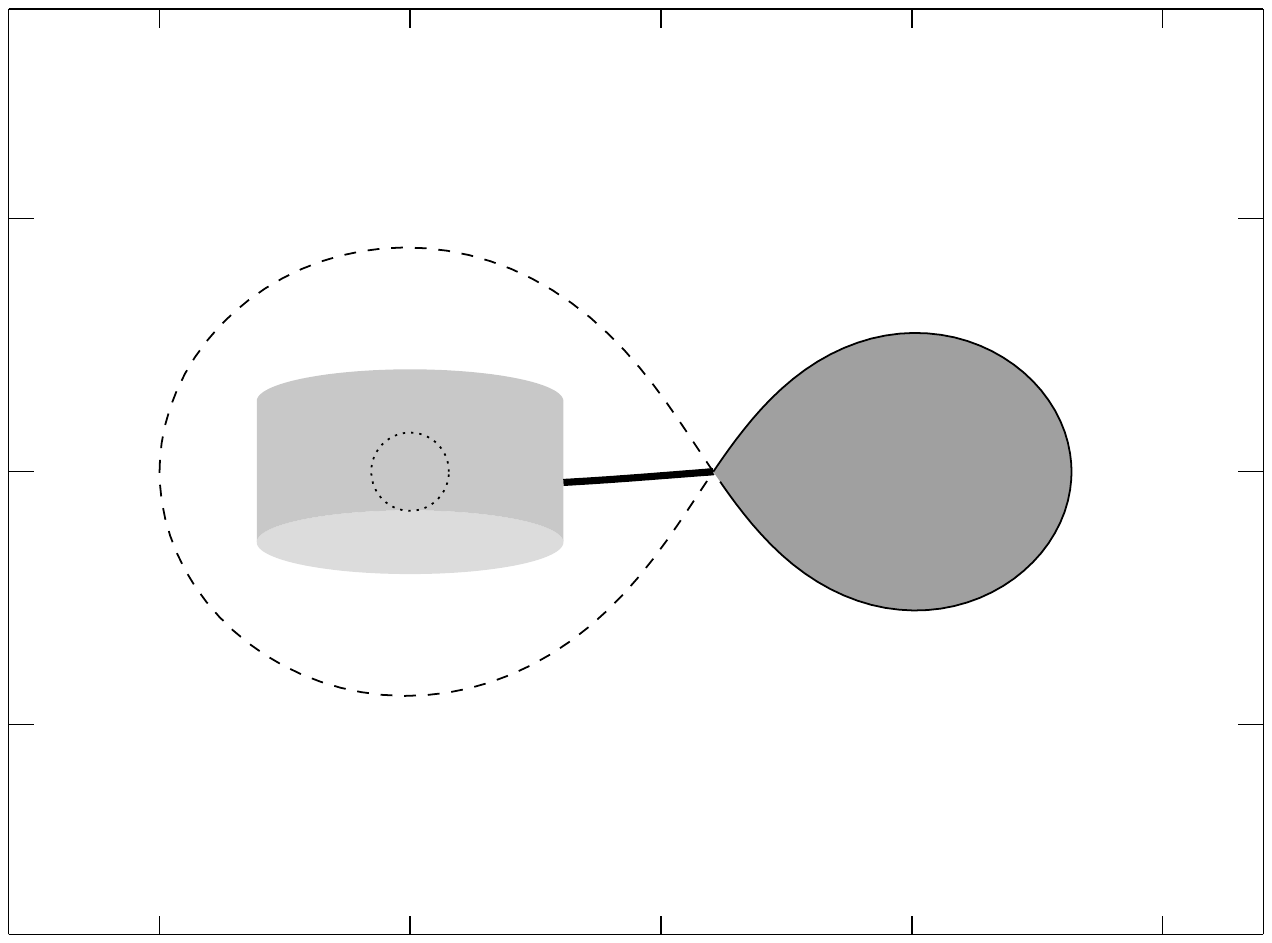}
	\vspace{-1.5cm} 
	\caption{{\it Above:} A schematic diagram of the components of W~Ser as seen from above the 
        orbital plane.  The Roche lobe-filling, mass donor star (right) launches a gas 
        stream (central dark curve) that enters the accretion torus (light grey disk) 
        surrounding the mass gainer (white circle at left).  The gas from the cool donor 
        star mixes in with the hotter torus gas, and this results in a slightly cooler temperature 
        in the region where the gas stream meets the torus.  The gas in the outer radial 
        parts of the torus gradually rises in temperature as the accreted gas is carried 
        around the outer disk.  The local temperatures assumed in the model are labeled in 
        the outer regions of the disk.  The dashed line indicates the Roche lobe boundary of
        the mass gainer, the plus sign shows the location of the center of mass, the arrows 
        show the orbital velocities of the stars, and the orbital phases for different 
        orientations are given in the periphery of the diagram. The spatial tick mark 
        interval corresponds to one half of the semimajor axis ($0.5\times 48.69 R_\odot = 24.34 R_\odot$).
        {\it Below:} The appearance of the binary model at orbital phase 0.25 for an inclination of 
        $i=101^\circ$. The dotted circle at left shows the location of the mass gainer star
        that is hidden by the vertical extent of the outer torus. }
        \label{fig:model}
\end{figure*}

We applied the same approach in fitting the TESS light curve, but we added a scheme to 
consider a temperature change at the torus rim that is related to azimuthal position. 
The incoming gas stream from the mass donor has a lower temperature than the average 
gas temperature in the outskirts of the torus, and this would tend to lower the torus gas 
temperature in the vicinity of the stream-torus intersection point.  As this accreted 
gas is carried around the torus, it would increase in temperature through mixing with 
the hotter torus gas.  Thus, we might expect that the average gas temperature at the 
torus rim would be slightly lower at positions just ahead of the axis joining the stars 
and slightly higher at azimuthal positions below the axis (Fig.~2).  Such a temperature 
variation could explain the observed lower flux before and higher flux after the eclipse
(Fig.~1).  Note that in some Algol binaries there is evidence of a heated region 
where the incoming gas stream strikes the accretion disk or gainer star \citep{Peters1984}. 
Indeed, \citet{Weiland1995} consider such a hot spot as the possible source of the 
\ion{Si}{4} $\lambda\lambda 1393, 1403$ emission lines in the spectrum of W~Ser.
However, such a hot spot would add to the observed flux at orbital phases before 
the eclipse, when in fact the system appears relatively faint. Thus,  
we suspect that any shock heating at the stream - torus meeting point is less evident in 
the case of W~Ser because the torus is large, vertically extended, and has a relatively 
low gas density compared to an equatorially confined accretion disk. 

We implemented this change by calculating light curves for two temperatures bracketing 
the default torus rim temperature of 8.0~kK.  We then determined the model light curve 
flux as a function of orbital phase by interpolating between these two models for an 
assumed relation between torus azimuthal position and temperature.  We set the temperature
variation function as an inverse sawtooth waveform centered on the eclipse that was 
smoothed with a Gaussian transfer function of FWHM = $51^\circ$.  This produced a smoothly
varying function with a maximum (minimum) near phase 0.1 (0.9).  The only fitting parameter
is the maximum temperature deviation from the default value.  We found we could make a 
good fit of the TESS light curve with a full temperature variation at the disk rim of 
$\pm 130$ degrees.  This range of rim temperature is labeled within the torus in Figure~2 
at the approximate positions of the temperature extrema and mean values.  The model 
light curve is shown as a gray line in Figure~1, and the introduction of the temperature 
variation in the torus rim is able to match the observed flux decline between eclipses.  

\newpage
\section{Fast Variations}
\label{sec:fast}

The residuals from the orbital fit of the light curve are shown in the lower part
of Figure~1, and they show clearly the faster than orbital fluctuations that were
so striking in earlier light curve investigations.   The periodic content of the 
variations is plotted in the power spectrum given in Figure~3.  This shows the 
Discrete Fourier Transform of the residuals corrected for aliases using the 
CLEAN algorithm \citep{Roberts1987}.  There is little power at high frequencies, 
and the main signals correspond to periods longer than 2 days.  The most prominent 
peak occurs at a frequency of $0.357 \pm 0.038$ cycles per day.  The corresponding 
periodic variation of $2.80 \pm 0.30$ days is shown as a dotted line superimposed on the 
residuals in Figure~1.  It appears that a quasi-sinusoidal variation is present 
that may last for six or so cycles before losing coherence. 

\begin{figure*}[h] 
        \centering
	\includegraphics[width=18cm]{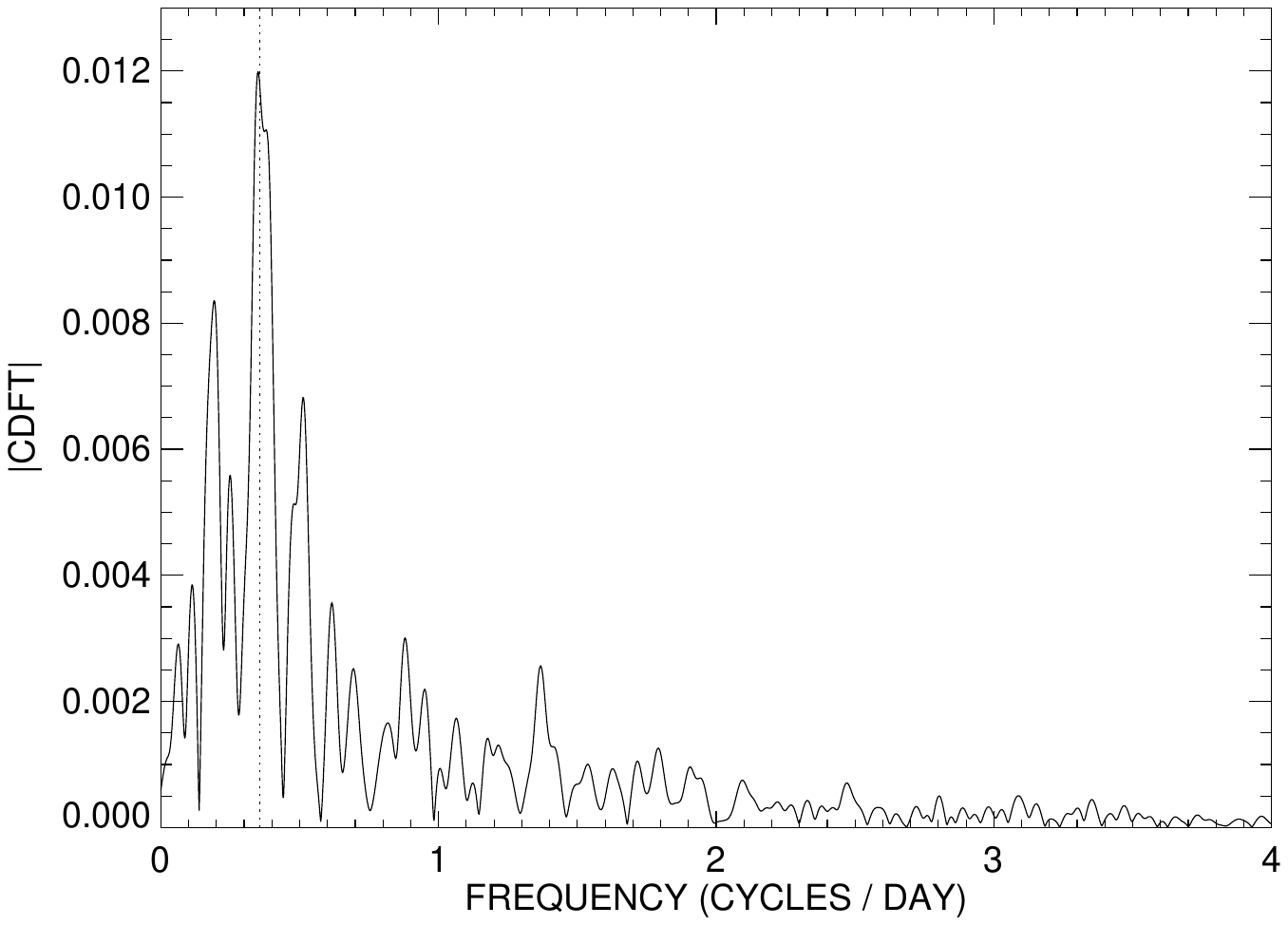}
	\vspace{-1.5cm} 
	\caption{The CLEANed Discrete Fourier Transform (CDFT) of the residuals from 
        the observed minus model light curve of W~Ser.  The vertical dotted line marks 
        the primary oscillatory peak at a frequency of 0.357 cycles per day.}
        \label{fig:cdft}
\end{figure*}

The main flux contributing component to the TESS measurements is the edge rim of the 
bright torus surrounding the gainer star (Fig.~2).  The orbital period for torus gas 
at the outer rim is 2.79 days for a gainer mass of $5.7 M_\odot$ \citep{Shepard2024}. 
This is the same within uncertainties as the fast variation seen in the TESS light curve. 
We suggest the fast variations are the result of turbulent knots in the torus gas
that create dense and bright sub-regions at the torus rim. 

\newpage

\section{The 1956 Light Curve}
\label{sec:fresa}

A comparison of the TESS light curve with that obtained by \citet{Fresa1957} in 1956 
provides us with a glimpse into changes that have occurred over timescales of decades. 
\citet{Fresa1957} made photoelectric observations on 109 nights over a period of 134 days, 
and because of the diurnal sampling, these measurements are useful to study variations 
on timescales longer than 2 days.  The photometry was obtained with an unfiltered 
IP21 photocell that was sensitive to flux in the 3000 -- 6000 \AA\ range.  His measurements 
are presented in Table~1 of his paper, and these are provided in a machine-readable table 
that is available with the electronic version of our paper.  


\citet{Fresa1957} estimated that the orbital period was $P=14.1563$ days in 1956. 
We rescaled our model light curve to this period, and we found that a small phase offset 
($+0.022$) was required to register the observations and model in orbital phase. 
The revised eclipse epoch for the 1956 observations is HJD 2,435,686.247.
The flux normalized version of the 1956 light curve is shown as a function of 
orbital phase in Figure~4.  The full temporal version of the light curve is given in 
Figure~1 in the paper by \citet{Fresa1957}.  The model light curve from the fit
of the TESS data is also shown in Figure~4.  The 1956 light curve shows similar 
eclipses and the general flux decline between eclipses as seen in the TESS light
curve, but the residuals from the fit show a larger dispersion, and there is a local 
flux maximum near orbital phase 0.5 that is absent in the TESS light curve. 

\begin{figure*}[h] 
        \centering
	\includegraphics[width=18cm]{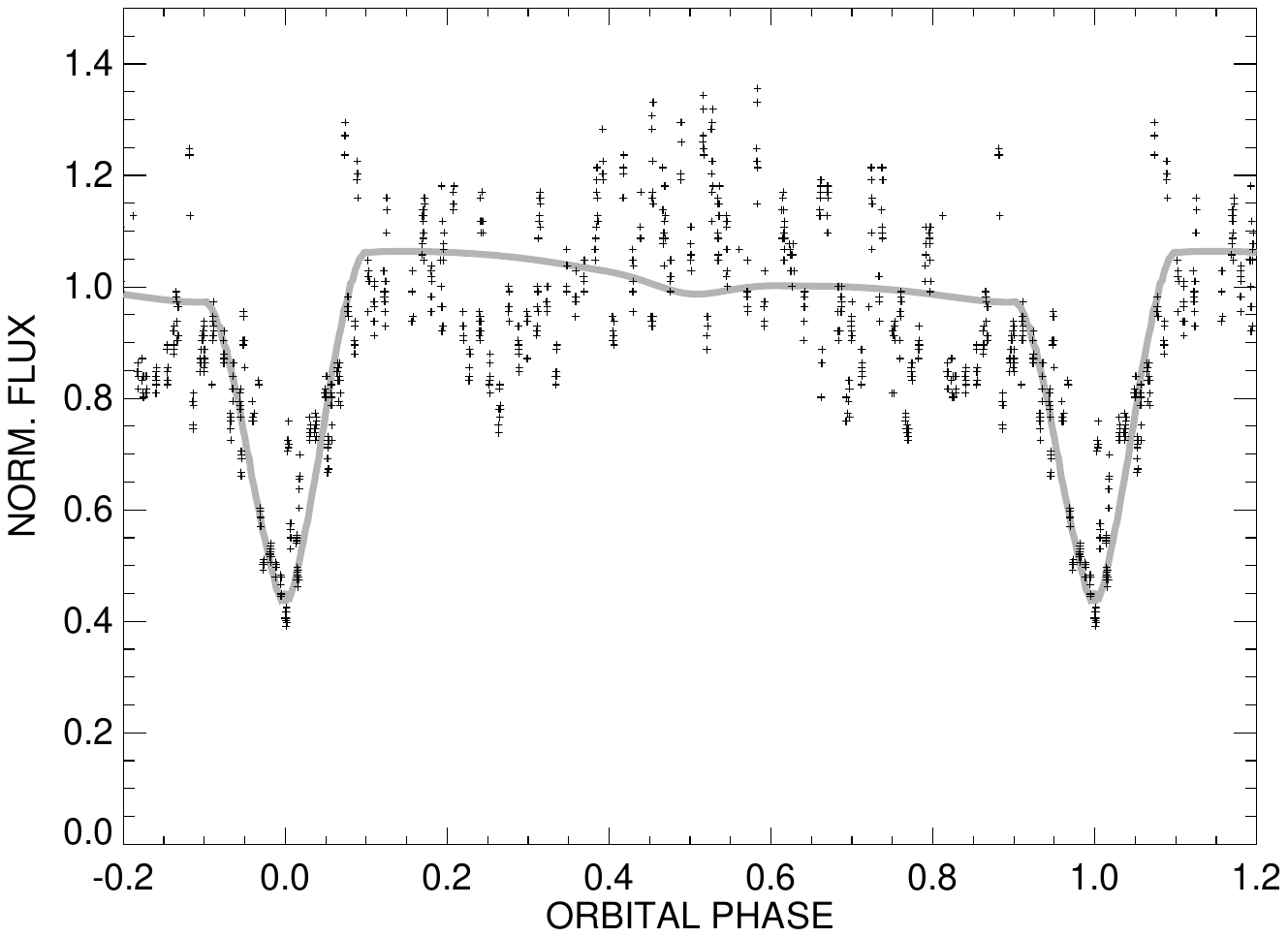}
	\vspace{-1.5cm} 
	\caption{The light curve of W~Ser obtained in 1956 by \citet{Fresa1957} plotted as a function of orbital phase.
	They grey-shaded line shows the model light curve from the TESS observations (Fig.~1).}
        \label{fig:flc}
\end{figure*}

\pagebreak

The power spectrum of the time series formed by subtracting the model from the observed
1956 light curve is presented in Figure~5.  The corresponding power spectrum for the TESS 
light curve (Fig.~3) is plotted as a dotted line.    The comparison shows that the same 
rapid signal found in the TESS data at a frequency of 0.357 cycles per day (Section 4) 
was also present in the 1956 light curve with approximately the same amplitude.  This is 
consistent with the visual impression of a 2.8 day oscillation that is apparent in some sections 
of the temporal light curve in Figure~1 of \citet{Fresa1957}.  There are other stronger 
peaks in the 1956 power spectrum on time scales related to the orbital period, half the 
orbital period, and the full observing season.  The orbital and half orbital peaks are related 
to the local brighter appearance in the light curve near orbital phase 0.5 (Fig.~4). 
The low frequency peak corresponds to a period of 117 days that 
is comparable to the timespan the observations (134 days).  This may be related to longer,
year-to-year variations that were found by \citet{McLaughlin1961} from observations obtained 
between 1928 and 1931.  The local maximum at phase 0.5 appearing in the 1956 light curve
is not evident in the mean adjusted light curve from 1928 -- 1930 (see Fig.~2 of 
\citealt{McLaughlin1961}), nor in the 2024 TESS light curve (Fig.~1).   Taken together, 
the photometric record indicates that significant changes occur beyond those on the orbital and 
2.8 day timescales highlighted in this paper.

\begin{figure*}[h] 
        \centering
	\includegraphics[width=18cm]{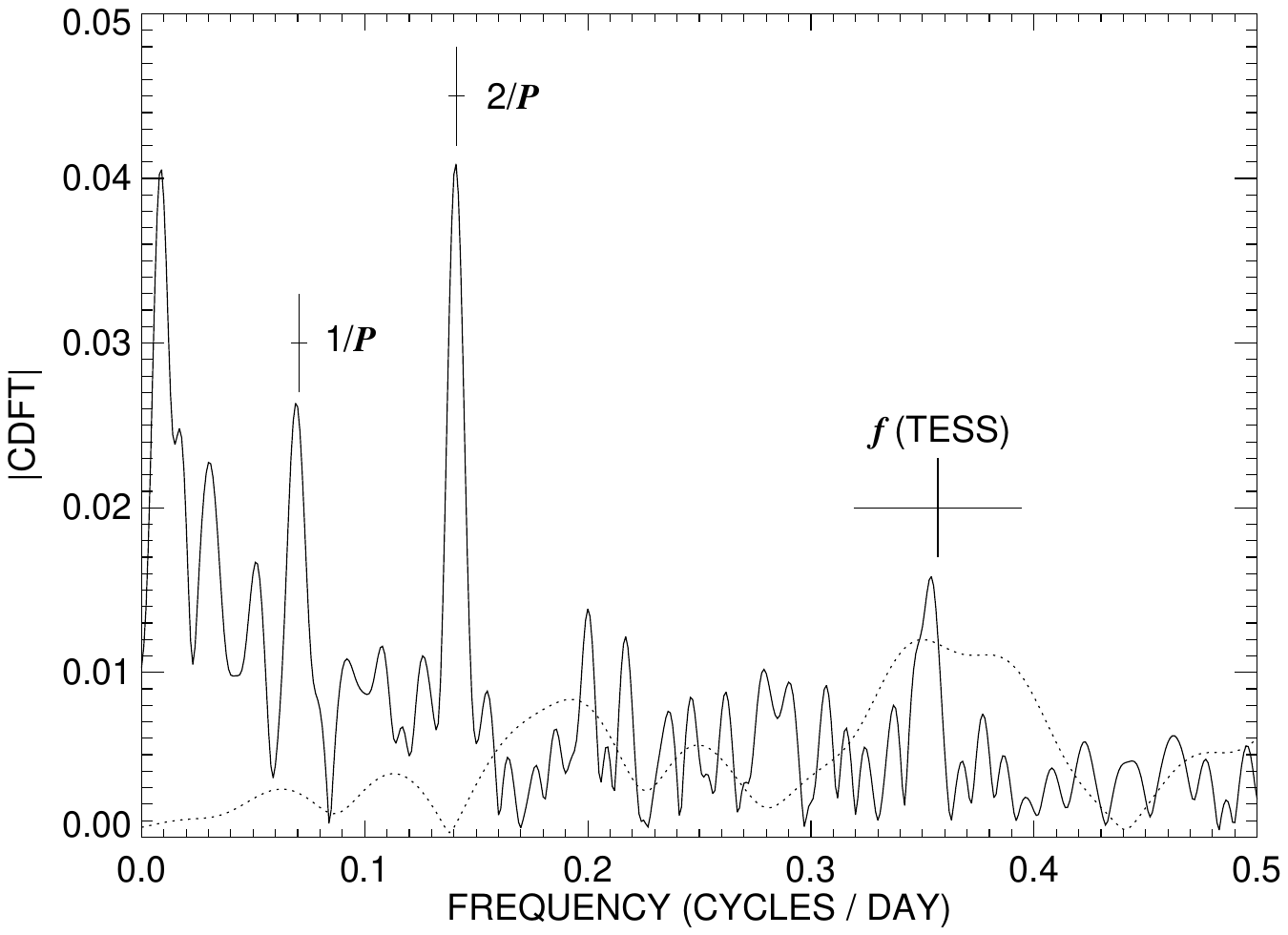}
	\vspace{-1.5cm} 
	\caption{The CLEANed Discrete Fourier Transform (CDFT) of the residuals from 
        the 1956 observed minus model light curve. The markers above the main peaks indicate the 
        frequency and uncertainty range of signals corresponding to the orbital period ($1/P$), 
        half the orbital period  ($2/P$), and the fast oscillation period found in the TESS light curve
        ($f ({\rm TESS})$). The dotted line shows the CDFT from the TESS data (Fig.~3).}
        \label{fig:fcdft}
\end{figure*}

\newpage
\section{Conclusions}
\label{sec:conclusions}

The exquisite TESS light curve of W~Ser has solved several of the mysteries about 
the system's unusual light curve.  The eclipse timings verify the period increase 
observed over the last century, and the period derivative implies that large scale 
mass transfer is taking place.  The gainer cannot easily accrete this mass flow, 
and the gas accumulates in a large accretion torus that effectively blocks the 
gainer from view.  The orbital variation in the TESS light curve is well matched 
by a model for this optically thick torus.  In particular, the strange jump in 
brightness across the eclipse can be explained by considering how cooler gas enters 
the torus at an azimuthal position slightly ahead of the direction of the donor star.  
The TESS light curve shows that the dominant timescale of the stochastic
or fast variations is the same as the gas orbital period at the outer rim of the torus. 
This same periodicity is present in the detailed 1956 light curve from \citet{Fresa1957}.
 These photometric variations are probably caused by relatively brighter 
 sub-regions in the outer rim of gas torus that are carried into and out of view as 
 those regions orbit the mass gainer star.   Thus, high quality, fast cadence 
 photometry offers us a window into the fast changing weather conditions in the 
 outer accretion torus.   We will soon 
have two more opportunities to follow such variations when TESS again captures the 
flux variations of W~Ser in future observing sectors 90 and 91 (in 2025 April and May). 

\begin{acknowledgments}
We are grateful to the referee for suggesting the comparison to the 1956 observations by \citet{Fresa1957}.
Funding for the TESS mission is provided by NASA’s Science Mission directorate.
Our work on W~Ser began with the support of National Science Foundation Grant Number AST-1411654.
Institutional support has been provided from the GSU College of Arts and Sciences and 
the GSU Office of the Vice President for Research and Economic Development.
NDR is grateful for support from NASA grants 80NSSC23K1049 and 80NSSC24K0229 
along with support from the Cottrell Scholar Award \#CS-CSA-2023-143 sponsored by 
the Research Corporation for Science Advancement.
\end{acknowledgments}

\begin{contribution}
DRG wrote the text and prepared the figures. 
KAS provided the background research on W~Ser. 
AK obtained and prepared the TESS light curve. 
NDR suggested the project and reviewed the manuscript.  
\end{contribution}

\facilities{TESS}

\software{ELC}

\bibliography{ms1}{}
\bibliographystyle{aasjournal}

\end{document}